\documentclass[preprint]{aastex}
\usepackage{gensymb}

\shorttitle{Ionization Fraction in DM Tau}
\shortauthors{\"Qi et al.}

\begin{document}

\title{The Ionization Fraction in the DM Tau Protoplanetary Disk}

\author{
Karin I. \"Oberg\altaffilmark{1}, Chunhua Qi,  David J. Wilner, Sean M. Andrews}
\affil{Harvard-Smithsonian Center for Astrophysics, 60 Garden Street, Cambridge, MA 02138, USA}

\altaffiltext{1}{Hubble Fellow}

\begin{abstract}

We present millimeter-wave observations of several molecular ions in 
the disk around the pre-main-sequence star DM~Tau and use these to 
investigate the ionization fraction in different regions of the disk.
New Submillimeter Array (SMA) observations of 
H$_2$D$^+$ J=$1_{1,0}-1_{1,1}$, N$_2$H$^+$ J=4--3 and CO J=3--2 are presented. H$_2$D$^+$ and N$_2$H$^+$ are not detected and using the CO 3--2 disk size the observations result in an upper limit of $<$0.47 K km s$^{-1}$ for both lines, a factor of 2.5 below previous single-dish H$_2$D$^+$ observations. 
Assuming LTE, a disk midplane temperature of 10--20~K and estimates 
of the H$_2$D$^+$ $o/p$ ratio, the observed limit corresponds to 
N$_{\rm H_2D^+}<4-21\times10^{12}$ cm$^{-2}$. 
We adopt a parametric model for the disk structure from the literature and use new 
IRAM 30 meter telescope observations of the H$^{13}$CO$^+$ J=3--2 line 
and previously published SMA observations of the N$_2$H$^+$ J=3--2, 
HCO$^+$ J=3--2 and DCO$^+$ J=3--2 lines to constrain the ionization 
fraction, $x_i$, in three temperature regions in the disk where 
theoretical considerations suggest different ions should dominate:
(1) a warm, upper layer with T$>$20~K where CO is in the gas-phase and 
HCO$^+$ is most abundant, where we estimate $x_i \simeq 4\times10^{-10}$, 
(2) a cooler molecular layer with T = 16--20~K where 
N$_2$H$^+$ and DCO$^+$ abundances are predicted to peak, with 
$ x_i \simeq3\times10^{-11}$, and (3) the cold, dense midplane with T$<$16~K 
where H$_3^+$ and its deuterated isotopologues are the main carriers of 
positive charge, with $x_i <3\times10^{-10}$.
While there are considerable uncertainties, these estimates are consistent with a decreasing ionization fraction into the deeper, colder, and denser disk 
layers.
Stronger constraints on the ionization fraction in the disk midplane
will require not only substantially more sensitive observations of 
the H$_2$D$^+$ $1_{1,0}-1_{1,1}$ line, but also robust determinations
of the $o/p$ ratio, observations of D$_2$H$^+$ and stronger constraints on where N$_2$ is present in the gas phase. 

\end{abstract}


\section{Introduction}

The disks around pre-main sequence stars are the sites of planet formation. 
Models of disk evolution depend on many disk properties, including the 
total disk mass, the detailed density and temperature structure, and 
accretion characteristics. Determining these properties are key to constrain 
how, when and where in disks planets can form. Since the main mass constituent 
in these disks is cold molecular hydrogen that cannot be observed directly, 
observations of dust and trace gas molecules are required to access the disk midplane 
properties. The ionization fraction $x_i$ in disks is especially important because the 
magneto-rotational instability (MRI), believed to drive viscous accretion, 
requires ionization to couple the magnetic field to the gas \citep{Gammie96}. 
Insufficient ionization may suppress the MRI and create a quiescent 
``dead zone'' with important implications for processes such as dust settling, 
planetesimal formation, and planet migration \citep{Ciesla07, Matsumura09}.

\begin{figure*}[htp]
\centering
  \includegraphics[angle=270,scale=0.5]{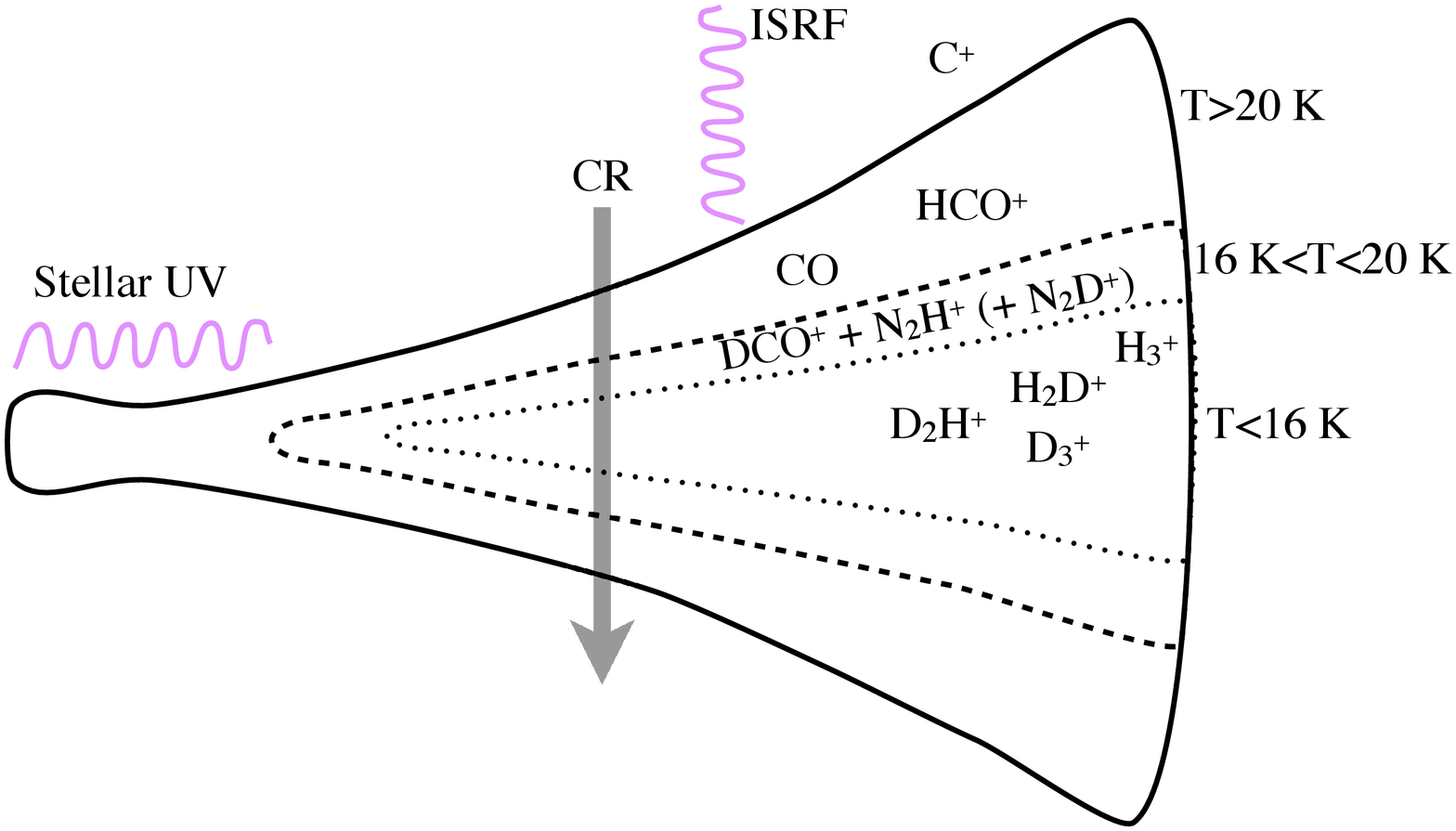}
\caption{
Schematic of the expected disk ion distribution as a function of disk layer temperature. The full lines represent the CO photodissociation front, the dashed lines the CO snowline, and the dotted line the transition region where the last heavy molecules freeze out.
\label{fig0}}
\end{figure*}

According to standard chemical models, disks are comprised of three
characteristic layers \citep[e.g.][]{Willacy98,Aikawa99}. Figure \ref{fig0} shows a schematic of the proposed layered structure focusing on ions.
The uppermost layer is exposed to ultraviolet radiation and X-rays that 
dissociate molecules and photoionize molecules and atoms to produce, e.g. 
C$^+$. Deeper into the disk, the radiation field is attenuated and molecules 
such as CO become abundant. In this cooler molecular layer, HCO$^+$ is 
predicted to be the most abundant ion \citep[e.g.][]{Aikawa06}.  
At even lower heights, at temperatures below 20~K, CO begins to freeze out onto grains. 
N$_2$ can remain in the gas-phase at slightly lower temperatures than CO 
\citep{Oberg05}, but in the deepest midplane layer, where temperatures drop 
below 16~K, the only molecules predicted to remain in the gas phase are 
H$_2$, H$_3^+$ and their isotopologues. Indeed, H$_2$D$^+$ and 
D$_2$H$^+$ may provide the best, and possibly the only, observable tracers 
of the gas kinematics and the ionization fraction in the cold, dense 
midplane \citep{AsensioRamos07}. It is important to note that in addition to the vertical layers, each region has an inner truncation radius where the midplane temperature exceeds the CO and N$_2$ freeze-out temperatures, respectively. Ions such as H$_2$D$^+$ are therefore only expected in the outer disk, beyond the CO snowline, whose radius will depend on the luminosity of the central star \citep[e.g.][]{Qi11}.

DM~Tau is a pre-main sequence star located $\sim$140~pc away from the Sun, 
surrounded by a gas-rich Keplerian circumstellar disk first discovered 
by \citet{Guilloteau94}. With a mass of 0.04~M$_\odot$, the disk is one 
of the more massive protoplanetary disks known \citep{Andrews11}. 
It is a transition disk with an imaged central hole of radius 19~AU 
\citep{Calvet05,Andrews11}. 
The DM~Tau disk shows strong emission from species proposed to trace
cold gas \citep{Oberg10c}, in particular DCO$^+$ and N$_2$H$^+$, and 
has previously been detected in a variety of molecular lines using 
single-dish telescopes \citep{Dutrey96}. 
It is therefore an especially good target to search for H$_2$D$^+$
emission. \citet{Ceccarelli04} reported a detection of the 
H$_2$D$^+$ $1_{1,0}-1_{1,1}$ line at 372 GHz at the 4.7~$\sigma$ level
toward DM Tau using the Caltech Submillimeter Observatory (CSO). 
However, \citet{Guilloteau06} suggest that the DM Tau result is 
``at best a 2$\sigma$ detection'' when analyzed with a proper model.
To help to resolve these conflicting claims, we have used the Submillimeter 
Array to observe this H$_2$D$^+$ line toward DM~Tau with improved 
sensitivity.  We analyze these results, together with observations of other 
molecular ions, to provide new constraints on the ionization fraction in 
different regions of the disk.

The paper is organized as follows. 
In \S2, we describe the SMA observations and complementary IRAM~30 meter 
telescope observations of molecular ions in the DM Tau disk. 
In \S3, we present upper limits on H$_2$D$^+$ and N$_2$H$^+$ J=4--3 
emission and detections of the CO J=3--2 and  H$^{13}$CO$^+$ J=3--2 lines and other resolved observations of
molecular ions from the literature.
In \S4, we use the line intensities together with an empirically derived disk 
structure to constrain the midplane temperature (\S4.1) and to calculate 
upper limits on the $o-$H$_2$D$^+$ column (\S4.2), the total H$_2$D$^+$ column 
(\S4.3), the total midplane ion column (\S4.4), the midplane ionization 
fraction (\S4.5), and the ionization fraction in higher disk layers (\S4.6). 
In \S5, we compare the results with predictions from a generic disk chemistry model.

\section{Observations\label{sec:obs}}

\subsection{SMA observations of CO, N$_2$H$^+$ and $o$-H$_2$D$^+$}

DM Tau was observed with the Submillimeter Array on top of Mauna Kea, Hawaii, 
on 2010 October 7. Six of the array antennas were available for these
observations, arranged in a compact configuration spanning baselines of 
16 to 77 meters. Using the dual receiver mode, we simultaneously targeted 
the $ortho$-H$_2$D$^+$ $1_{1,0}-1_{1,1}$ line at 372.4213~GHz and 
the N$_2$H$^+$ J=4--3 at 372.6725~GHz with the ``high frequency'' receiver,
and the 345.796 GHz CO J=3--2 line with the ``low frequency'' receiver.
Table~1 summarizes the spectral line setup. At these frequencies, 
the synthesized beam sizes were $2\farcs2\times1\farcs7$ and 
$2\farcs4\times2\farcs0$, respectively. 

The observing conditions were generally very good, with 
atmospheric opacity $\tau_{\rm 225{\:GHz}}\sim0.04$ measured at the 
nearby Caltech Submillimeter Observatory, and stable atmospheric phase. 
This opacity corresponds to a zenith transmission of $\sim50$\% at 
the H$_2$D$^+$ line frequency.
Short observations of the two calibrators 3C120 and J0449+113 
were interleaved with the observations of DM~Tau for gain calibration.
The bandpass response was calibrated with observations of Uranus and the 
available bright quasars 3C273, 3C454.3 and 3C84. Observations of Uranus 
and Callisto provided the absolute scale for flux densities, via comparison to theoretical models for their emission, and the two calibrators agree within the expected uncertainties, but all reported data were calibrated with Callisto.
The systematic uncertainty in the absolute flux scale is $\sim$10\%. 
The data were edited and calibrated with the IDL-based MIR software 
package\footnote{http://www.cfa.harvard.edu/$\sim$cqi/mircook.html}.
Continuum and spectral line images were generated and CLEANed using MIRIAD. 
To check that the flux calibration is accurate, we compared the continuum 
fluxes of 0.22~Jy at 0.87~mm and 0.30~Jy at 0.81~mm with previously published 
values \citep{Andrews05,Andrews07} and found that they agree within the 
reported uncertainties. 

\subsection{IRAM 30m observations of H$^{13}$CO$^+$}

The H$^{13}$CO$^+$ J=3--2 line at 260.255~GHz was observed toward 
DM~Tau on 2011 January 25 with the IRAM 30~meter telescope. 
The observations were carried out with the EMIR 330 GHz receiver, with a 
beam size of $\sim$9\farcs5 (FWHM). The receiver was connected to a unit of 
the autocorrelator with a spectral resolution of 320~kHz and a bandwidth of 
240~MHz, equivalent to an unsmoothed velocity resolution of 
$\sim$0.4 km s$^{-1}$. Typical system temperatures were 300-400 K. 
The observations were carried out using wobbler switching with a $100''$ throw. 
Pointing was checked every $\sim$2 hours on J0430+052 and J0316+413 with a 
typical accuracy of $<2''$. The main-beam brightness temperature was 
calculated from the antenna temperatures in {\it pako} using reported 
main beam and forward efficiencies (B$_{\rm eff}$ and F$_{\rm eff}$) of 
88\% and 53\%, respectively. The data were reduced with the CLASS program, 
part of the GILDAS software package\footnote{http://www.iram.fr/IRAMFR/GILDAS}. 
Linear baselines were determined from velocity ranges without emission 
features, and then subtracted from the spectra. 

\section{Results}\label{sec:res}

\begin{figure*}[htp]
\centering
\epsscale{1}
\plotone{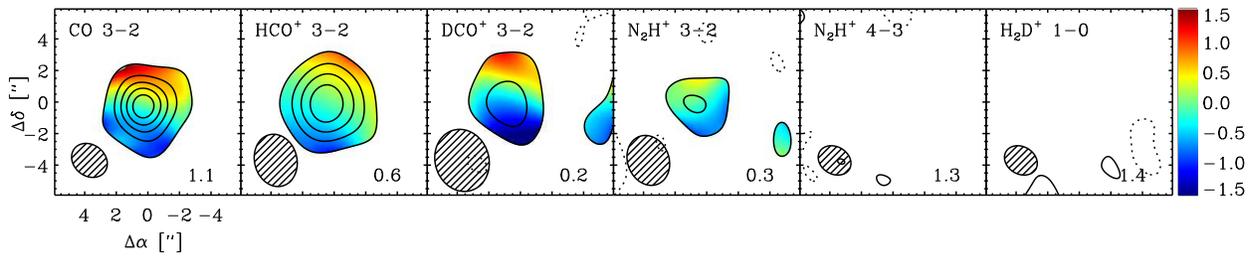}
\caption{
Moment maps of the CO J=3--2 line and the molecular ion lines toward DM Tau
observed with the SMA. 
The contour level (in Jy beam$^{-1}$) are shown in the bottom right corner
of each panel and correspond to 5$\sigma$ for CO and HCO$^+$ and 2$\sigma$ 
for the other species.
The color scale indicates the velocity field around the velocity centroid 
of 6.1 km/s. 
The synthesized beams are plotted in the lower left hand corners of each panel.
\label{fig1}}
\end{figure*}

We did not detect either the H$_2$D$^+$ $1_{1,0}-1_{1,1}$ line or 
the N$_2$H$^+$ J=4--3 line. 
Figure \ref{fig1} shows imaging upper limits for these lines, along with 
moment maps for the CO J=3--2 line and three lines from other molecular ions,
HCO$^+$ J=3--2, DCO$^+$ J=3--2 and N$_2$H$^+$ J=3--2 \citep{Oberg10c}. 
The velocity gradient due to disk rotation appears similar in all of the 
moment maps, even for the weak N$_2$H$^+$ and DCO$^+$ emission.
The apparent disk sizes are also similar within the uncertainties. 
To estimate the disk emission region, we fit an elliptical Gaussian to the 
CO J=3--2 visibilities. The resulting major and minor FWHM are 
3\farcs1 and 2\farcs5. The effective disk size is defined as twice the FWHM of the CO emission, which will contain 98\% of the emission assuming a Gaussian distribution. This provides a convenient definition of the disk size, which is not model dependent, but it is important to note that the defined disk radius does not directly correspond to the outer edge of the disk when the disk is modeled as a power law, or to the critical radius when the disk is modeled with the similarity solution.
With the adopted definition, the CO 3--2 disk is 6\farcs2$\times$5\farcs0, corresponding to a disk radius of $\sim$400~AU, which compares well with previous gas size estimates \citep{Guilloteau94}. The FWHM of all ions are consistent with the CO disk within the fit uncertainties. The synthesized beam sizes across the disk major axis vary between 2\farcs0 and 4\farcs0, corresponding to spatial scales of 280--560~AU. The observations therefore only resolve the chemistry in the outer disk.

\begin{figure}[htp]
\centering
\epsscale{0.5}
\plotone{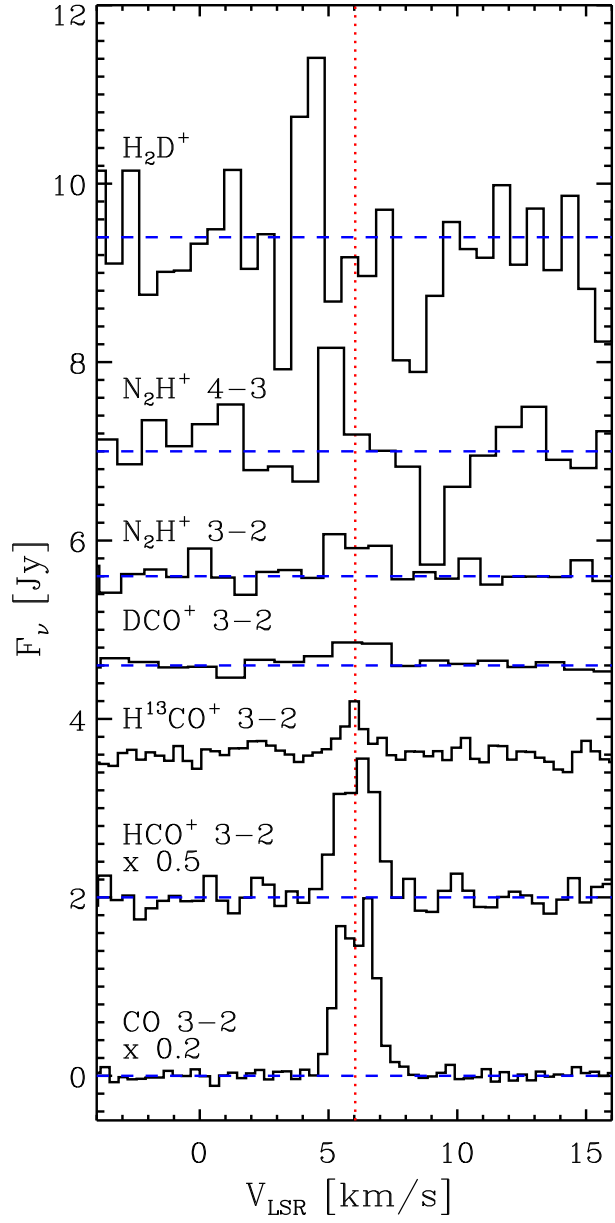}
\caption{
Spatially integrated spectra of CO J=3--2 and all searched for ion lines 
toward the DM Tau disk, including non-detections of the N$_2$H$^+$ J=4--3 
and H$_2$D$^+$ $1_{1,0}-1_{1,1}$ lines. 
The HCO$^+$ and CO lines have been scaled down for visibility.
\label{fig2}}
\end{figure}

Figure \ref{fig2} shows the spatially integrated spectra of the species 
imaged in Figure \ref{fig1}. The integration is carried out using an elliptical integration area the size of the CO disk.
Table \ref{tab2} lists the line fluxes integrated over the velocity range from 2 to 10 km~s$^{-1}$. 
The HCO$^+$ J=3--2 has been previously observed toward DM~Tau using the 
IRAM~30m telescope by \citet{Dutrey97}, and the integrated intensity of 
4.1[0.5]~Jy~km~s$^{-1}$ compares well with the SMA value of
5.3[0.4]~Jy~km~s$^{-1}$ \citep{Oberg10c}, considering the mutual calibration uncertainties, which are in addition to the rms in brackets.

To convert the SMA fluxes to intensities, we adopt the CO J=3-2 size scale 
for all of the emission lines and use the flux-to-intensity conversion 
formula: $T_{\rm b} = F\times13.6\times(\lambda)^2 / (a\times b)$, 
where the constant factor is specific for the SMA observations, $T_{\rm b}$ is the intensity in K, $F$ the flux per beam (here the 6\farcs2$\times$5\farcs0 CO disk) in Jy, $\lambda$ the wavelength of the line in millimeters and 
$a$ and $b$ are the major and minor axes of the beam (disk) in arc seconds (Table \ref{tab2}). 
For the H$_2$D$^+$ and N$_2$H$^+$ J=4-3 lines, we calculate 2$\sigma$ upper 
limits with the standard formula: 
$\sigma=rms\times FWHM / \sqrt{n_{\rm ch}}$, where $n_{\rm ch}$ is 
the number of channels across the CO 3--2 line at FWHM.  
The $2\sigma$-limit on the H$_2$D$^+$ line intensity is $<$0.47 K km s$^{-1}$. 
Averaged over the 22$\arcsec$ beam of the Caltech Submillimeter Observatory,
this 2$\sigma$ upper limit corresponds to $<$0.03 K km s$^{-1}$, a factor of 2.5 below the claimed detection by 
\citet{Ceccarelli04}. This conclusively rules out the previous detection claim and entails that H$_2$D$^+$ is not yet detected in any protoplanetary disk. This is consistent with the upper limits found by Chapillon et al. (subm. to A\&A) using the JCMT. 

Figure \ref{fig2} also includes the detection of H$^{13}$CO$^+$ J=3--2 from 
the IRAM 30 meter telescope. The reported intensity in Table \ref{tab2} has been scaled to account 
for beam dilution by $9.5^2/(6.2\times5.0)$, i.e. the ratio of the IRAM~30 
meter beam area over the CO disk area. The fact that the H$^{13}$CO$^+$ line 
is detected indicates that the main isotopologue is optically thick, since if the emission ratio of HCO$^+ / $ H$^{13}$CO$^+>30$, the H$^{13}$CO$^+$ line would have been buried in the noise.

\section{The Disk Ionization Fraction}

We use the observations to constrain the ionization fraction in the disk 
in a four-step process.
First, the observed upper limit on the 
$o$-H$_2$D$^+$ 1$_{\rm1,0}$-1$_{\rm1,1}$ line is used to constrain the 
$o$-H$_2$D$^+$ column density N($o$-H$_2$D$^+$) for a range of plausible temperatures for the
disk midplane.
Second, we use recent model results on the temperature dependent $o/p$ ratio
to convert N($o$-H$_2$D$^+$) limits to total ($o+p$) 
H$_2$D$^+$ limits, i.e. N(H$_2$D$^+$). 
Third, we use calculations of the H$_3^+$/H$_2$D$^+$/D$_2$H$^+$/D$_3^+$ 
ratios for the appropriate range of temperatures to constrain the total 
column density N($\sum$H$_{3-x}$D$_x$) of these ions in the midplane. 
Finally, we use the total column density of ions in the midplane together
with a physical model of the DM Tau disk and measurements of N(HCO$^+$), N(DCO$^+$), and N(N$_2$H$^+$) to constrain $x_i$ in the disk midplane, an intermediate layer where CO has started
to freeze out, and the upper warm molecular layer. 

Before embarking on this series of steps, we place observational constraints 
on the temperatures in the disk midplane with the aim to limit the range of temperatures to consider when deriving a midplane $x_i$ from the $o$-H$_2$D$^+$ 1$_{\rm1,0}$-1$_{\rm1,1}$. For this investigation, we define the 
midplane as the region where all heavy elements are frozen out and H$_3^+$ 
and its isotopologues are the dominant charge carriers. Theoretically, the 
midplane temperature is limited by the freeze-out temperatures of CO and 
N$_2$, since both react readily with H$_3^+$, resulting in HCO$^+$ and 
N$_2$H$^+$ \citep{AsensioRamos07}. Since N$_2$ is maintained in the gas phase 
at slightly lower temperatures than CO \citep{Oberg05}, the N$_2$H$^+$ rotational temperature provides 
the strongest observational limit available on the midplane temperature. 

\subsection{Limits on the midplane temperature from 
N$_2$H$^+$ 1--0, 3--2 and 4--3
\label{sec:midplane_temp}}

To estimate the N$_2$H$^+$ rotational temperature, we assume LTE and 
co-spatial emission from the N$_2$H$^+$ 3--2 and 4--3 lines.
This is a reasonable approximation because
chemical models predict that N$_2$H$^+$ should be abundant in a narrow 
temperature regime between where CO begins to freeze-out and 
N$_2$ freeze-out is not complete \citep{Bergin06, Dutrey07}. Assuming optically thin emission, unity filling factor within the 
CO 3--2 disk, and molecules thermalized at a single rotational temperature, 

\begin{equation}
N = \frac{1.67\times10^{14}}{\nu\mu^2S} Q(T_{\rm rot})e^{\rm E_u/T_{rot}}\int T_{\rm b}dv
\end{equation}

\noindent from \citet{Thi04}, where $N$ is the total number of molecules, 
$Q(T_{\rm rot})$ is the temperature dependent partition function, 
$E_{\rm u}$ the energy of the upper level in K, 
and $T_{\rm rot}$ the rotational temperature. 
We obtain the partition function $Q(T_{\rm rot})$ by second order interpolation
of the listed values in CDMS\footnote{The Cologne Database for Molecular Spectroscopy at http://www.astro.uni-koeln.de/cdms/catalog}, $\mu^2S$ is gathered directly, and $E_{\rm u}$ 
is calculated from the listed $E_{\rm low}$ values: $E_{\rm u}=E_{\rm low}+0.0334\times\nu$, where $\nu$ is the line frequency in GHz. Then an upper limit on the N$_2$H$^+$ temperature can be calculated from

\begin{eqnarray}
T_{rot} = \frac{E_u^{4-3}-E_u^{3-2}}{ln\left(\frac{\nu_{4-3}\mu_{4-3}^2S_{4-3}}{\nu_{3-2}\mu_{3-2}^2S_{3-2}}\frac{\int T_{\rm b}^{3-2}dv}{\int T_{\rm b}^{4-3}dv}\right)}.
\end{eqnarray}

\noindent Using the integrated intensity of the J=3--2 line and the 2$\sigma$ 
upper limit on the J=4--3 line, we obtain a 2$\sigma$ upper limit on the 
N$_2$H$^+$ rotational temperature of $<$26~K. This is consistent with the 
predictions that N$_2$H$^+$ should become abundant below the CO freeze-out 
temperature of 20~K, but does not provide any stronger constraints. 
It is also consistent with the N$_2$H$^+$ J=1--0 line reported by 
\citet{Dutrey07}. We estimate the line flux from their Figure~1 to be 
$<$0.1 Jy~km~s$^{-1}$. 
Using this limit and the integrated intensity of the J=3--2 line results in 
a lower limit on the N$_2$H$^+$ rotational temperature of 11~K. 
More sensitive observations of the N$_2$H$^+$ lines would be valuable to 
constrain this excitation temperature further. In the absence of more 
detailed information, we calculate LTE column densities of midplane ions for 
a range of temperatures, from 10 to 20~K. The 20~K boundary is set by laboratory experiments on CO and N$_2$ freeze-out -- above this temperature a majority of CO and N$_2$ are maintained in the gas-phase quickly destroying any formed H$_3^+$ \citep{Bisschop06}. The CO snowline was recently reported to correspond to a freeze-out temperature of  19~K, consistent with predictions \citep{Qi11}.

\subsection{N($o$-H$_2$D$^+$)}

In general, column densities are calculated most accurately by constructing 
a self-consistent chemical-physical disk model and applying a radiative 
transfer code to predict line emission profiles that can be compared directly 
to observations. Without a detection of the H$_2$D$^+$ line, however, 
it is difficult to constrain where the H$_2$D$^+$ emission arises in the disk.  
We therefore opt for a classical LTE approach 
\citep[e.g.][]{vanDishoeck03}, 
calculated for the range of reasonable disk midplane temperatures 
(see \S\ref{sec:midplane_temp})
to put limits on the disk averaged N($o$-H$_2$D$^+$). 

A significant complication for the LTE calculation is that $ortho$ ($o$) and 
$para$ ($p$) H$_2$D$^+$ can behave as separate species at low temperatures 
\citep{Flower04,Sipila10} resulting 
in a non-thermal partitioning between the $o$ and $p$ states. 
To calculate the partition function for $o$-H$_2$D$^+$, we use a two-level 
approximation, which assumes no transitions between the $o$ and $p$ H$_2$D$^+$,
and that transitions between the ground state and first excited state dominate
within each ``species'' of H$_2$D$^+$.
This approximation is expected to be valid at temperatures of 10 to 20~K, 
since E$_{\rm u}$ is 104~K for the first excited state of $o$-H$_2$D$^+$.
The $o$-H$_2$D$^+$ partition function is calculated from

\begin{equation}
Q=g_u\times e^{-E_u/T_{\rm ex}}+g_l\times e^{-E_l/T_{\rm ex}}
\end{equation}

\noindent 
where the degeneracies, $g_u$ and $g_l$, are equal to 3. 

Assuming LTE (Equation 1), 
excitation temperatures in the range 10 to 20~K and the 
partition function appropriate for these temperatures, N($o$-H$_2$D$^+$) limits are calculated from the observed upper limit on the 
H$_2$D$^+$ J=1$_{\rm1,0}$-1$_{\rm1,1}$ line flux. 
As shown in Figure 3a, the limits on the disk average N($o$-H$_2$D$^+$) 
range from $1.3-2.7\times10^{12}$ cm$^{-2}$ (with the highest value 
corresponding to the lowest temperature).

\subsection{Total ($o + p$) N(H$_2$D$^+$)}

The limit on the total N(H$_2$D$^+$), i.e. $o + p$, depends 
on the $o$-H$_2$D$^+$ column density and the temperature dependent $o/p$ ratio. 
This ratio has been modeled by \citet{Sipila10} from 4 to 20~K for a 
molecular hydrogen density of 10$^6$ cm$^{-3}$ and an interstellar grain 
distribution. The derived $o/p$ ratio is 0.5--0.1 for temperatures of 
10--20 K, with the lowest ratio at 13--17~K. 
While the assumed grain size distribution and density are more appropriate for 
dense clouds than disk midplanes, the grain size distribution mainly affects 
the temperature profile, and \citet{Sipila10} find that the H$_2$D$^+$ $o/p$ 
is nearly constant with increasing density. Thus we expect the values 
calculated with these assumptions should be valid for disk midplane conditions.

Figure~3b shows the result of applying the literature $o/p$ ratios to the 
previously calculated $o$-H$_2$D$^+$ column density limits to obtain limits 
on N(H$_2$D$^+$) as a function of temperature. 
These limits range from $4$ to $21\times10^{12}$ cm$^{-2}$.
The presence of a peak value at 13~K is a direct consequence of the fact
that the limit on N($o$-H$_2$D$^+$) decreases with temperature, 
while the $o/p$ ratio has a minimum at 13--17~K.

\subsection{Total Midplane Ion column density: N($\sum$ H$_{3-x}$D$_x^+$)}

The limit on the total column density of ions in the midplane, N($\sum$~H$_{3-x}$D$_x^+$), 
depends on N(H$_2$D$^+$) and the ratio of 
N(H$_2$D$^+$) to N($\sum$ H$_{3-x}$D$_x^+$). 
Like the $o/p$ ratio, this ratio is expected to vary with density 
and temperature, as well as other environmental properties such as 
grain size \citep{Flower04,Ceccarelli05,Sipila10}. 
This ratio will also depend on whether or not CO and N$_2$ are frozen out 
completely, or if low abundances of these species can be maintained in the 
gas-phase through non-thermal desorption \citep{AsensioRamos07}.
\citet{Caselli08} modeled the ratios of all deuterated H$_3^+$ isotopologues 
for molecular hydrogen density 10$^5$ cm$^{-3}$ and interstellar grains as 
a function of temperature. As mentioned previously, the disk midplane density 
is likely higher than assumed in these models. \citet{Sipila10} found 
 an increasing importance of D$_3^+$ with increasing density at 10~K, but it is unclear whether this effect is important also at higher temperatures. Another potential unknown 
is the abundance of H$^+$, though \citet{Ceccarelli05} found that its 
contribution is negligible at the high densities of the disk midplane. 

With these caveats in mind, we use the ratios derived by \citet{Caselli08} 
with maximum depletion, i.e. minimum amounts of CO and N$_2$ in the gas-phase, 
to calculate upper limits on N($\sum$ H$_{3-x}$D$_x^+$) in the disk 
midplane as a function of assumed temperature. 
Figure 3c shows the results, with limits that range from $1.5$ to 
$8.5\times10^{13}$ cm$^{-2}$, with the peak value at 13~K. The modification 
of the shape compared to the H$_2$D$^+$ curve is due to the fact that the 
N(H$_2$D$^+$)/N($\sum$ H$_{3-x}$D$_x^+$) ratio peaks near 13~K and then falls off quickly at higher temperatures.

\subsection{Ionization Fraction ($x_i$) in the Midplane}

To derive limits on the midplane ion abundances, we adopt the model of
the DM~Tau disk density and temperature structure described by 
\citet{Andrews11} based on parametric fitting of the spectral energy 
distribution and resolved observations of millimeter-wave dust emission. This model does not include any chemistry, but rather provides an observationally constrained physical structure, which can be used to derive the gas masses in different temperature layers in the disk. This is needed to convert an ion column density into an ion abundance.
We apply a standard interstellar gas-to-dust ratio of 100 to convert the 
dust densities in this model to gas densities. To determine a midplane mass, 
we define the disk midplane as all material $\lesssim$16~K, the 
temperature for complete N$_2$ freeze-out in most astrophysical environments based on laboratory 
measurements \citep{Oberg05,Bisschop06}, i.e. the region where it is most 
likely that H$_3^+$ and its isotopologues become abundant. In this disk layer, 
the density weighted temperature is 11~K, and the disk averaged gas column 
density N$_{\rm H_2}$ is $15\times10^{22}$ cm$^{-2}$. Using this column
density,  and the upper limit on the ion column density in the midplane of $4\times10^{13}$ cm$^{-2}$ from Table 3 and  Fig. 4, $x_i<2.7\times10^{-10}$ n$_{\rm H}$. 
Alternatively, if the onset of CO freeze-out at 20~K defines the region where H$_2$D$^+$ is abundant rather than 
N$_2$ freeze-out at 16~K, then the density averaged ion temperature 
is 13~K, and $x_i<4.5\times10^{-10}$ n$_{\rm H}$.
Table~\ref{tab3} summarizes the midplane ($<$16~K) ionization results and also shows how the ionization fractions compare if H$_2$D$^+$ is only assumed to be present in the midplane or in the entire T$<$20~K region. 
It is important to keep in mind that all of these calculations are based 
on model dependent values, and better limits on the midplane ionization 
abundance requires not only more sensitive $o-$H$_2$D$^+$ observations, 
but also direct constraints on the $o/p$ ratio and the relative abundances
of different isotopologues. In particular the latter estimate may be off by an order of magnitude from different model predictions of the importance of D$_3^+$ compared to H$_2$D$^+$.

\subsection{Ionization in different disk layers}

It is interesting to compare the estimates of ionization fraction in different 
layers of the disk, since this may impact e.g. MRI activation and accretion 
flow.  The mass of each layer, defined by a temperature interval, can be estimated from the same DM Tau physical model as was used in \S4.5. Between 20~K, where CO begins to freeze out, and the CO photodissociation region, HCO$^+$ is predicted to be 
the most abundant ion \citep[e.g.][]{Willacy07}. Below 20~K, as CO starts to 
freeze out, N$_2$H$^+$ becomes more abundant, since reactions with CO 
are the main destruction pathway of N$_2$H$^+$. Below 16~K, CO and N$_2$ are 
completely frozen out and  H$_3^+$ and its isotopologues become the main charge carriers. 

The DCO$^+$ abundance peaks between the region where HCO$^+$ dominates and the 
midplane \citep{Willacy07, Aikawa06} and therefore is assumed to coexist with 
N$_2$H$^+$ in the intermediate layer. In this layer, DCO$^+$/HCO$^+$ exceeds 
unity and the HCO$^+$ contribution to the charge balance should be small. This model result is supported by comparing DCO$^+$ line intensities: \citet{Guilloteau06} detected DCO$^+$ 2--1 at $0.29\pm0.03$~Jy km s$^{-1}$ and the 
intensity ratio of the 3--2/2--1 lines implies a rotational temperature of 
11--23~K, assuming LTE.

For the midplane and the low-temperature intermediate layer, the emission regions of the ions 
are expected to be well confined, and the assumption of LTE at a single 
rotational temperature should be a reasonable first approximation. 
For the warm molecular layer where HCO$^+$ dominates, the assumption
of LTE is not {\it a priori} a good approximation since this layer spans a 
temperature range from 20 to 100~K or higher. However, the derived HCO$^+$ 
column density is not very sensitive within this range; it varies by less than a factor of three when using Eq. 1 over the full range of plausible temperatures and  the LTE approach retains utility. As discussed below, the outcome of this 
LTE calculation compares well with a radiative transfer calculation of
HCO$^+$ emission in the DM~Tau disk \citep{Pietu07}.

We estimate the disk averaged N(HCO$^+$), N(N$_2$H$^+$) and N(DCO$^+$) using Equation 1, the observed intensities from Table \ref{tab2}, and the mass weighted temperatures from the DM Tau structural model of 38~K for the CO dominated layer ($>$20~K) and 18~K for the intermediate layer where N(N$_2$H$^+$) and N(DCO$^+$)  are most abundant (16--20~K).
Table 3 lists the results and Figure \ref{fig4} shows the resulting column densities together with the H$_3^+$ limit (LTE at 11~K), within their respective temperature regimes ($<$16~K, 16--20~K, and $>$20~K), corresponding to the different disk layers. 
Note that N(HCO$^+$) is calculated from H$^{13}$CO$^+$ assuming a standard $^{12}$C/$^{13}$C 
ratio of 70, since the high opacity of the main isotopologue underestimates N(HCO$^+$) by a factor of seven. 

The calculation results in a disk averaged DCO$^+$/HCO$^+$ ratio of $\sim$0.05. This is comparable to a previous estimate of 0.035 toward the disk around 
TW Hya \citep{Thi04}. 
Both exceed the cosmic D/H ratio of $\sim$10$^{-5}$ by three orders of 
magnitude, demonstrating an efficient deuterium fractionation in these disks.
The derived ratio is an order of magnitude above the ratio found by 
\citet{Guilloteau06}, who used resolved HCO$^+$ 1-0 and single-dish DCO$^+$ 2-1
emission and a radiative transfer calculation rather than LTE. 
The derived disk averaged N(HCO$^+$) are almost identical over the 
inner 400~AU of the disk: $7\times10^{12}$ cm$^{-2}$ from their tabulated 
column density at 300~AU and power law index, compared to 
$9\times10^{12}$ cm$^{-2}$ calculated here assuming LTE at 38~K. 
The different ratios must thus be due to different assumptions on the DCO$^+$ 
emission conditions. The HCO$^+$ column density compares well with model predictions of the outer disk (100--400~AU). It is for example consistent with the disk chemistry model by \citet{Willacy07} within a factor of 3.

\citet{Dutrey07} observed N$_2$H$^+$ 1--0 with the Plateau de Bure 
interferometer 
and derived N$_2$H$^+$/HCO$^+$ of $\sim$0.02--0.03. This ratio agrees well with the results in Table 3, which implies 
a disk-averaged ratio of 0.03.

The disk averaged N(H$_2$) within each temperature regime are 
15, 2.4, and $2.6\times10^{22}$ cm$^{-2}$, and these values are used with the derived ion column densities to
calculate disk averaged $x_i$ in each layer. 
Figure \ref{fig4} shows that the HCO$^+$ abundance and the midplane 
ion limits are comparable at $\sim3-4\times10^{-10}$ n$_{\rm H}$, while 
the abundance of DCO$^+$~+~N$_2$H$^+$ is an order of magnitude lower at 
$3\times10^{-11}$ n$_{\rm H}$. 

Overall, the derived ion abundances suggest a decreasing $x_i$ 
with depth in the disk, which might be expected from attenuation of ionizing 
radiation. The assumption that DCO$^+$ and N$_2$H$^+$ alone are representative for the degree of ionization in the intermediate layer may be oversimplified, however. The $16<T<20$ layer is a transition region where some HCO$^+$ may still be present and H$_3^+$ may already have started to become important. 
A combination of better limits on the midplane ions as well as more detailed 
modeling of the intermediate layer with DCO$^+$ and N$_2$H$^+$ will be 
needed to confirm the depth dependent trend on $x_i$.

\section{Discussion} \label{sec:disc}

In the previous section the disk averaged column densities and abundances were calculated based on observed line intensities and a density and temperature structure of DM Tau from the literature. This particular structure model has not been used as a basis for chemical modeling. In this section we therefore use predictions on the ion chemistry in generic disk models to compare our observational results with the current understanding of the chemical compositions in disks.

Model predictions on the abundances of H$_2$D$^+$ and other ions in 
protoplanetary disks have been made by many authors 
\citep[e.g.][]{Aikawa06, Willacy07, AsensioRamos07,Fogel11} for generic T Tauri disk structures.  
In particular, the models of \citet{Willacy07} include abundances and 
column densities for all of the ions considered in this study, and therefore 
we focus on comparing the observational results with these predictions. 
We examine specifically 
the disk averaged ionization fraction, 
the ionization fraction in different disk layers, 
and the relative abundances of different ions. 

\citet{Willacy07} based their chemical calculations on the disk structure 
model of \citet{dAlessio01}, representing a generic T Tauri disk. The assumed accretion rate, disk mass etc are within a factor of a few of the values that have been derived for DM Tau  \citep{Calvet05}. Most disk chemistry predictions are therefore expected to be valid for DM Tau, except for absolute ion column densities.  The chemical modeling is based on the UMIST database, updated with multiple deuterated species \citep{Rodgers96}. 
Within this model framework, \citet{Willacy07} investigated the importance 
of cosmic ray desorption and photodesorption on the disk chemical abundances,
which has a large effect on the ion abundance ratios.  

An important caveat when comparing observations and model predictions, is that \citet{Willacy07} presents column densities as a function of radius with no simple procedure to calculate disk averaged column densities. Observations can therefore only be compared with model predictions at specific radii. From the DM Tau structure model the observed emission is dominated by the outer disk; beyond 50 AU the column densities of the different temperature layers are almost constant, which entails that the larger emission region of the outer disk will dominate. In particular more than 2/3 of the ion emission is expected to originate outside of 250~AU if the ions are present across the CO disk. In the models by \citet{Willacy07} the ion column densities are constant within a factor of 2 at 100--400~AU, except for the relative importance of different H$_3^+$ isotopologues. It is therefore useful to compare model predictions in this regime with the observed disk averaged values as long as this factor of two uncertainty is considered. \citet{Willacy07} provides the most complete model results at 250~AU and this radius is used in all comparisons between model and observations. 

\subsection{Disk averaged ionization fraction}

The main sources of ionizing radiation in disks are ultraviolet (UV) photons, 
cosmic rays, radioactive decay and X-rays \citep[e.g.][]{Willacy07}, and the 
ionization level therefore depends on a combination of the incident fluxes 
and the attenuation of external photons with dust and gas column density. 
\citet{Willacy07} include both the interstellar radiation field and a stellar 
UV field that is 500 times stronger at 100~AU. Because of efficient absorption 
by dust and H$_2$, the UV ionization is mainly important in the disk atmosphere
and upper molecular layer. 
Cosmic rays are assumed to penetrate unperturbed to the disk midplane 
resulting in an ionization rate of $1.3 \times10^{-17}$ s$^{-1}$.
Radioactive decay also contributes to the midplane ionization, at a rate
of $0.6 \times10^{-17}$ s$^{-1}$. The effect of X-rays are not included in the 
model and the model may therefore underestimate $x_i$ in lower disk layers. 

In this model, the total ion column density 
(H$_3^+$, H$_2$D$^+$, D$_2$H$^+$, D$_3^+$, HCO$^+$, DCO$^+$, N$_2$H$^+$) 
at 250~AU is $2-3\times10^{13}$ cm$^{-2}$, corresponding to an average $x_i$ of $2-3\times10^{-10}$ n$_{\rm H}$. This value is consistent with the observed abundances and limits. In the related disk model of \citet{AsensioRamos07}, the ion abundance 
depends on the assumed cosmic ray flux, as expected. 

 \subsection{Vertical ionization gradients}
 
 In disks, including the model from \citet{Willacy07} the temperature increases both inward and upward in the disk, resulting in a vertically layered structure outside of an inner truncation radius. In the model, HCO$^+$ extends through the disk, while DCO$^+$ is only abundant outside of 50~AU and H$_2$D$^+$ outside of 100~AU. As mentioned above, beyond 100~AU the different ion column densities are basically constant, justifying the comparison between disk averaged observations and model predictions at 250~AU.
 
Because all UV photons and probably some of the X-rays and cosmic rays are 
absorbed in the outer disk layers, the ionization fraction should decrease 
between the outer disk layers and the disk midplane. 
At 250~AU, in the disk chemistry models \cite[Fig. 3 in][]{Willacy07} the ion abundance in the upper molecular 
layer, where CO is in the gas-phase and HCO$^+$ is the dominant ion, 
is a factor of 5 higher than in the midplane 
($\sim5\times10^{-10}$ compared to $\sim1\times10^{-10}$ n$_{\rm H}$). 
In the intermediate region, where DCO$^+$ and N$_2$H$^+$ abundances peak, 
the ion abundance is about a factor of 2--3 below the CO layer ion abundance. This is qualitatively consistent with the 
results of the analysis of the DM Tau observations (see Figure~4). 

\subsection{Specific ion column densities, abundances and ion ratios}

{\it HCO$^+$ and DCO$^+$:} 
The HCO$^+$ column density predicted by \citet{Willacy07} 
is $\sim3\times10^{12}$ cm$^{-2}$ at 250~AU and is found to not 
depend on the different desorption mechanisms considered. 
In contrast, the DCO$^+$ column density predictions vary by an order 
magnitude, resulting in a DCO$^+$/HCO$^+$ ratio of 0.1--1. 
The higher DCO$^+$/HCO$^+$ ratios appear with desorption through cosmic ray 
heating, which maintains CO in the gas-phase in the disk midplane where 
the temperatures are low enough for efficient deuterium fractionation and 
enhanced DCO$^+$ production. 
The observations indicate that the disk averaged DCO$^+$/HCO$^+$ abundance 
ratio is below 0.1, which suggests that CO is confined to the warm 
$>$20~K molecular layer. 
The observational results are consistent with 
including UV photodesorption, which is efficient only in the warm molecular layer.

{\it N$_2$H$^+$:} 
The N$_2$H$^+$/HCO$^+$ abundance ratio is predicted to be in the range
from $2$ to $9\times10^{-3}$, but the models with the highest values also 
predict DCO$^+$/HCO$^+$ ratios of unity and can be excluded.
Compared to the value of 0.03 derived from the observations, the models are 
an order of magnitude too low. The reason for this discrepancy is unknown, demonstrating that
even the simplest nitrogen-based chemistry remains poorly constrained.

{\it H$_2$D$^+$:} 
N(H$_2$D$^+$) is predicted to be 10$^{12}$ cm$^{-2}$, with midplane abundance $\sim$10$^{-11}$ n$_{\rm H}$. 
These values are well below the observational limits,
and the reason is that D$_3^+$ is the dominant carrier of charge in the
midplane in these models, 20--50 times more abundant than H$_2$D$^+$. 
In the disk model by \citet{Aikawa06}, the H$_2$D$^+$ column density 
is predicted to be significantly higher, 10$^{13}$ cm$^{-2}$, likely
because multiple deuteration is not considered. Both models are
thus consistent with the current limits, and better constraints on the chemical 
makeup of the midplane will require not only more sensitive observations
of H$_2$D$^+$, but also robust $o/p$ determinations, and additional 
constraints on the abundance ratios of the isotopolgues. 

\section{Conclusions}

\begin{enumerate}
\item A new SMA observation of the H$_2$D$^+$ $1_{1,0}-1_{1,1}$ line
toward the DM Tau disk
results in a disk-averaged intensity upper limit of $<$0.47 K km s$^{-1}$.
\item Assuming LTE and a temperature dependent $o/p$ conversion factor, 
the observation corresponds to an H$_2$D$^+$ column density 
$<$4--21$\times10^{12}$ cm$^{-2}$, for midplane temperatures of 10--20~K.
\item Based on SMA and IRAM 30 meter telescope observations 
of additional molecular ions toward DM Tau and a previously published model 
of the disk structure, we estimate ionization fractions of $x_i=4\times10^{-10}$  
in the molecular layer where T$>$20~K (based on H$^{13}$CO$^+$), $x_i>3\times10^{-11}$ in the 16--20~K layer where 
N$_2$H$^+$ and DCO$^+$ are common, and $x_i<3\times10^{-10}$ in the midplane below 16~K where H$_3^+$ isotopologues dominate. The ion emission is expected to be dominated by the outer disk, beyond 250~AU, because of the larger emission area.
\item The estimated ionization fractions depend on adopted model predictions of $o/p$ ratios, deuteration levels and the dominating ions in different disk layers. Observational constraints on all three parameters together with deeper H$_2$D$^+$ observations are therefore key to calculating more accurate ionization fractions in disks.
\end{enumerate}

{\it Facilities:} \facility{SMA}

\acknowledgments

\noindent The manuscript has benefitted from comments by an anonymous referee and from discussions with Eric Herbst and Ted Bergin. The SMA is a joint project between the Smithsonian Astrophysical Observatory and the Academia Sinica Institute of Astronomy and Astrophysics and is funded by the Smithsonian Institution and the Academia Sinica. Support for K.~I.~O. is provided by NASA through a Hubble Fellowship grant  awarded by the Space Telescope Science Institute, which is operated by the Association of Universities for Research in Astronomy, Inc., for NASA, under contract NAS 5-26555. We also acknowledge NASA Origins of Solar Systems grant No. NNX11AK63 

\bibliographystyle{aa}

\newpage

\begin{figure}[htp]
\centering
\epsscale{0.5}
\plotone{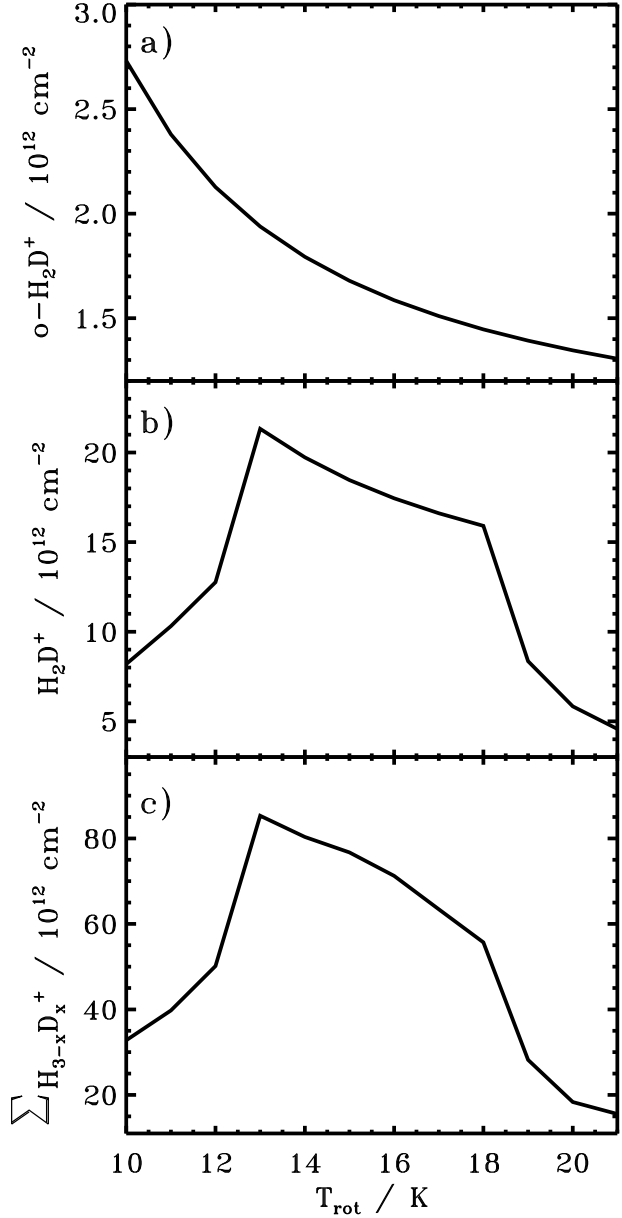}
\caption{
Panel a) shows disk averaged N($o$-H$_2$D$^+$) limits as a
function of the assumed rotational temperature. 
Panel b) shows the limits on N(H$_2$D$^+$), 
where the temperature dependence is due to a combination of the calculated 
N($o$-H$_2$D$^+$) limits and literature values on the 
temperature dependent $o/p$ ratio. 
Panel c) shows the column density limits on N($\sum$ H$_{3-x}$D$_x^+$), which includes the temperature dependent relative abundances 
of the different isotopologues (as well as the previous temperature 
dependencies).
\label{fig3}}
\end{figure}

\begin{figure}[htp]
\centering
\epsscale{0.5}
\plotone{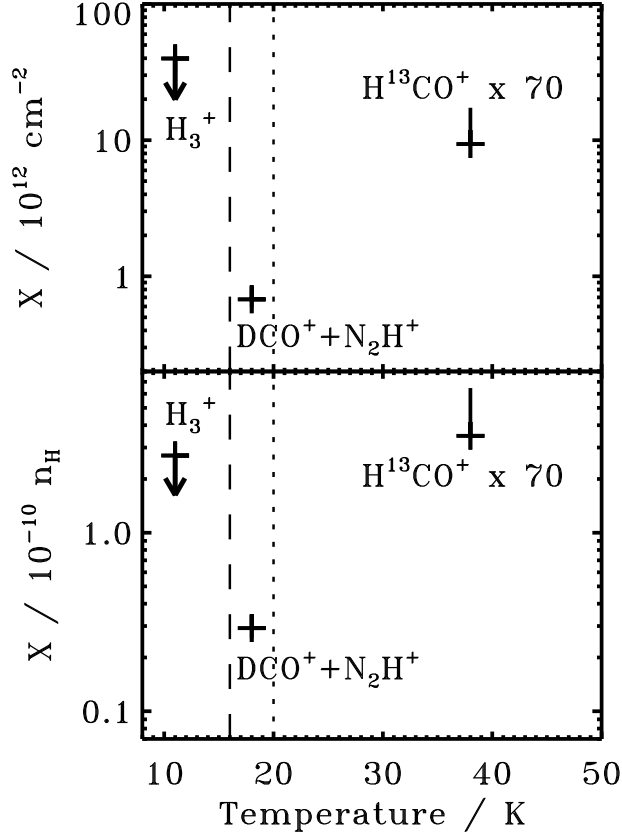}
\caption{Disk averaged ion column densities (upper panel) 
and ion abundances (lower panel) calculated at the density averaged 
temperature in the disk layers where 
H$_3^+$, DCO$^+$~$+$~N$_2$H$^+$ and HCO$^+$ (here traced by H$^{13}$CO$^+$) are predicted in chemical
models to carry most of the positive charge. 
The dashed line marks the transition between the midplane (where N$_2$ freezes out) and 
N$_2$H$^+$ layer, and the dotted line marks the CO freeze-out temperature. 
The combined uncertainties from line fluxes and rotational temperature estimates are generally smaller than the symbols, except for H$^{13}$CO$^+$ where the excitation temperature range is larger. 
\label{fig4}}
\end{figure}

\newpage

\begin{deluxetable}{lccccccc}
\tabletypesize{\footnotesize}
\tablecaption{\small SMA target spectral lines. \label{tbl:setup1}}
\tablewidth{0pt}
\tablehead{
\colhead{Chunk } & \colhead{Frequency range} & \colhead{Channels} & \colhead{Resolution}& \colhead{Lines} & \colhead{Frequency}& \colhead{T$_{\rm up}$}\\
 & (GHz) & & (km s$^{-1}$)& &(GHz)&(K)
}
\startdata
USB:S08 &345.75--345.85&256&0.36&CO 3--2 &345.7960&33\\
USB:S08 &372.37--372.48&128&0.66&H$_2$D$^+$ 1$_{\rm1,0}$-1$_{\rm1,1}$&372.4213&104\\
USB:S11 &372.63--372.73&128&0.66&N$_2$H$^+$ 4--3&372.6725&45\\
\enddata
\end{deluxetable}

\begin{deluxetable}{lcccc}
\tablecolumns{5} 
\tabletypesize{\footnotesize}
\tablecaption{Line data with 2$\sigma$ upper limits and 1$\sigma$ uncertainties and continuum fluxes.
\label{tab2}}
\tablewidth{0pt}
\tablehead{
\colhead{Line} & \colhead{$\int F dv$ / Jy km s$^{-1}$}& \colhead{$\lambda$ / mm}&\colhead{$\int I dv$ / K km s$^{-1}$} 
& \colhead{$F_{\rm cont}$ / mJy}}
\startdata
CO 3--2			&16.7[0.2]	&0.87&	5.43[0.06]		&216[22]\\
HCO$^+$ 3--2		&5.3[0.4]	&1.12&	2.9[0.2]		&143[14]\\
DCO$^+$ 3--2		&0.7[0.1]	&1.39&	0.62[0.08]		&104[10]\\
N$_2$H$^+$ 3--2	&1.0[0.1]	&1.07&	0.52[0.06]		&143[14]\\
N$_2$H$^+$ 4--3	&$<$1.8	&0.81&	$<$0.47		&303[30]\\
\smallskip
H$_2$D$^+$ 1$_{\rm1,0}$-1$_{\rm1,1}$	&$<$1.8	&0.81&	$<$0.47		&303[30]\\
H$^{13}$CO$^+$ 3--2&0.70[0.08]	&1.15&0.22[0.03]$^{\rm a}$\\
\enddata
\\$^{\rm a}$ After correction for beam dilution.
\end{deluxetable}

\begin{deluxetable}{lcccc}
\tablecolumns{5} 
\tabletypesize{\footnotesize}
\tablecaption{LTE ion column densities and abundances in different disk layers.
\label{tab3}}
\tablewidth{0pt}
\tablehead{
\colhead{} & \colhead{T Layer [K]} &\colhead{T$_{\rm rot}$ [K]} & \colhead{N$_{\rm x}$ [10$^{12}$ cm$^{-2}$] } & \colhead{n$_{\rm x}$ [10$^{-10}$ n$_{\rm H}$]}
}
\startdata
$o$-H$_2$D$^+$	&$<$16/$<$20 &11$^{\rm a}$ / 13$^{\rm b}$&$<$2.4$^{\rm a}$ / $<$1.9$^{\rm b}$&$<$0.08$^{\rm a}$ / $<$0.06$^{\rm b}$\\
H$_2$D$^+$		&$<$16/$<$20&11 / 13&$<$10 / $<$21	&$<$0.35	/ $<$0.6\\
\smallskip
$\sum$H$_{x}$D$_{3-x}^+$	&$<$16/$<$20&11 / 13 &$<$40 / $<$85 &$<$2.7 / $<$4.5\\
N$_2$H$^+$		&16--20&18	&0.31$\pm0.03$	&$0.14\pm0.02$\\
DCO$^+$			&16--20&18	&0.36$\pm0.05$	&$0.16\pm0.02$\\
HCO$^+$			&$>$20&38	&9.4$\pm0.8$		&$3.5\pm0.3$\\
\enddata
\\$^{\rm a}$ Density weighted temperature in the midplane layer when T$<$16~K.\\
$^{\rm b}$ Density weighted temperature in the midplane layer when T$<$20~K.
\end{deluxetable}

\end{document}